\documentclass[12pt]{article}
\usepackage{upgreek}
\usepackage{mathrsfs}
\usepackage{amsmath,amssymb,color}
\usepackage{amsfonts}
\usepackage[all]{xy}
\usepackage{url}
\usepackage{xcolor}
\usepackage[colorlinks,linkcolor=black,anchorcolor=blue,citecolor=blue,urlcolor=blue]{hyperref}
\usepackage{graphicx}
\usepackage{wrapfig}
\usepackage{cases}
\usepackage{appendix}
\usepackage[T1]{fontenc}
\usepackage[utf8]{inputenc}
\usepackage{authblk}
\usepackage{txfonts}
\usepackage[numbers,sort&compress]{natbib}
\newcommand{\ads}{{\rm AdS}}
\newcommand{\adst}{{\rm AdS}_3}

\newcommand{\adscft}{{\rm AdS/CFT}}

\newcommand{\cf}{\textrm{Crofton's formula}}

\newcommand{\er}[1]{\eqref{#1}}

\newcommand{\ci}[1]{}

\newcommand{\ke}{\rangle}
\newcommand{\br}{\langle}
\newcommand{\lb}{\left(}
\newcommand{\rb}{\right)}

\newcommand{\lsb}{\left[}
\newcommand{\rsb}{\right]}

\newcommand{\p}{\partial}

\newcommand{\ba}{\begin{eqnarray}}
\newcommand{\ea}{\end{eqnarray}}
\newcommand{\be}{\begin{equation}}
\newcommand{\ee}{\end{equation}}
\newcommand{\bal}{\begin{align}}
\newcommand{\eal}{\end{align}}
\newcommand{\bay}[1]{\left(\begin{array}{#1}}
\newcommand{\eay}{\end{array}\right)}
\newcommand{\eg}{\textrm{e.g.} }
\newcommand{\ie}{\textrm{i.e.}, }

\newcommand{\st}[1]{|#1\ke}

\newcommand{\zt}[1]{\textrm{#1}}

\def\rmd{{\rm d}}
\def\xa{{\alpha}}

\def\xb{{\beta}}

\def\xd{{\delta}}

\def\xe{{\epsilon}}

\def\xg{{\gamma}}

\def\xk{{\kappa}}

\def\xO{{\Omega}}

\def\xS{{\Sigma}}
\def\xt{{\theta}}

\def\CK{{\cal K}}

\def\BE{\mathbb{E}}

\def\BH{\mathbb{H}}

\def\h{{\frac 1 2}}

\title{Holographic Complexity from the Crofton's Formula in Lorentzian $\adst$}
\author[a,b]{Xing Huang \thanks{xingavatar@gmail.com}}
\author[a,b]{Le Zhang \thanks{leeviackerman@outlook.com}}
\affil[a]{Institute of Modern Physics, Northwest University, Xi'an 710069, China}
\affil[b]{Shaanxi Key Laboratory for Theoretical Physics Frontiers, Xi'an 710069, China}

\date{} 
\begin{document}

\maketitle
\begin{abstract}
We study the Crofton's formula in the Lorentzian AdS$_3$ and find that the area of a generic space-like two dimensional surface is given by the flux of space-like geodesics. The ``complexity=volume'' conjecture then implies a new holographic representation of complexity in terms of the number of geodesics. Finally, we explore the possible explanation of this result from the standpoint of information theory.
\end{abstract}

\newpage
\noindent\rule[0.1\baselineskip]{\textwidth}{0.8pt}
\tableofcontents
\noindent\rule[0.1\baselineskip]{\textwidth}{0.8pt}
\linespread{1.5}\selectfont

\section{Introduction}
As a standout of the holographic principle, the $\adscft$ correspondence \cite{maldacena} offers possibly the best playground for exploring quantum gravity as it indicates that the latter (likely to be string theory) in the Anti-de Sitter spacetime is equivalent to some well-defined conformal field theory on the boundary. It has long been conjectured that the bulk geometry shall emerge from the boundary entanglement as both should be universal \ie insensitive to
the discrepancies between the CFTs that admit gravity duals. A strong supporting evidence comes from the Ryu-Takayanagi formula \cite{ryu,ryu2}, which associates the area of a minimal surface anchored on the boundary with the entanglement entropy of some subregion in the boundary CFT.

Motivated by the Ryu-Takayanagi formula, Balasubramanian, et al. \cite{diffentropy} connected the length of a closed bulk curve on the time slice of $\adst$ to the differential entropy, which thus provides entropic interpretation for more general geometric objects. This has later been incorporated by Czech, et al. \cite{czech1} into the program of integral geometry, in which the length of a curve follows from the number of geodesics it has intersections with, which is known as the Crofton's formula in the mathematical literature \cite{santal}.

Complexity is another concept whose dual may help to uncover the information theoretic origin of gravitational physics. It is defined as the number of quantum gates (unitary operators) to produce a given state from a reference state. We can imagine the system to be a collection of qubits each located on a site of a spatial lattice. We would like the quantum gates to act only on a small number of neighboring qubits for our interests. Such restriction of locality is physically reasonable (as interactions are generally local particularly in the field theories we consider) and it is crucial to the definition of complexity. Accordingly the reference state is usually the one with no pre-existing long range correlation. For example, we can choose a state with all qubits set into the state $\st{0}$. The quantum gates form a quantum circuit \ie a unitary operator and hence it is natural to assign the complexity to the circuit itself. Such operator/circuit complexity can be defined in a geometric way pioneered by Nielsen and collaborators \cite{nielsen2006geometric, nielsen2006quantum, dowling2008geometry} (see also \cite{jefferson2017circuit} for recent developments), as the length of a geodesic in the space of all the unitary operators on the Hilbert space, measured by the cost function. Complexity provides ordering to the states in the Hilbert space. In this sense, it is argued that complexity can be understood as an analog of entropy (see \eg \cite{Susskind:2018pmk}). Naturally complexity can be also used as a probe of the chaotic behavior (see \eg \cite{Ali:2019zcj}).

It is difficult to define complexity in a field theory. We will instead work with those theories with holographic duals and focus on the holographic dual of complexity. There are two proposals on such duality that could turn out to be equivalent. Complexity may either be dual to the volume of the spatial slice (so-called complexity=volume, {\it a.k.a.} CV \cite{susskind2016computational}) or the gravity action inside the Wheeler-DeWitt patch (complexity=action, {\it a.k.a.} CA \cite{brown2016holographic, brown2016complexity}).

On the boundary CFT side, complexity can be defined using the Liouville action \cite{Miyaji:2016mxg, Caputa:2017urj, Caputa:2017yrh,Takayanagi:2018pml}. Recently the correspondence between a quantum circuit and a co-dimension one surface was proposed in \cite{Takayanagi:2018pml}. More precisely, the quantum circuit is realized as a path integral on the co-dimension one bulk surface $M[\xS(0)]$ whose boundary $\p M = \xS(0)$ is the constant time surface (taken to be $t = 0$) on the boundary. The evolution of the Euclidean time gives a one parameter family of quantum states, which correspond (in the sense of surface/state correspondence \cite{Miyaji:2015yva}) to the codimension-two surfaces $\xS(t)$ on $M[\xS(0)]$. This process is essentially the renormalization group flow and $t$ serves as the energy scale. Different surfaces $M[\xS(0)]$ lead to different induced metrics in the path integral, which can be written in the canonical form of $e^{2\phi} \eta_{\mu\nu}$. Due to conformal symmetry, all these path integrals are supposed to provide the same state with different normalization factors $e^{C(M[\xS(0)])}$ given by the Liouville action $C(M[\xS(0)]) = C_L(e^{2\phi} \eta_{\mu\nu})$ (the so-called PI complexity), which can be taken as an alternate definition of complexity (see \cite{caputa2019quantum,Camargo:2019isp} for the equivalence with the circuit complexity in Nielsen's approach). \footnote{In fact it might be possible to go one-step further to think that the bulk region enclosed by the surface is determined entirely by the path integral, which seems quite natural from the standpoint of surface/state correspondence even though the precise dictionary is unclear.} The CV conjecture then implies the areas of these surfaces are equal to the values of the cost function for the circuits, with the complexity being the minimum given by the optimized one \footnote{It should be noted that the area does not necessarily agree with PI complexity. In our opinion it is perfectly acceptable to have different definitions for complexity as long as they can be good enough approximation of each other. For example, the number of gates might be only polynomial in the circuit complexity defined geometrically \cite{nielsen2006quantum}. However, such generalization of CV does not seem to work in Lorentzian AdS as the extremal surface used to define complexity is the maximal one.}. Both the cost function and the path integral complexity can be defined for a generic circuit and we will loosely refer to them as the cost of a circuit.

It is known \cite{santal,huang} that the area of a codimension-one surface on a time slice or more generally in any Euclidean AdS-like space can be computed using the flux of geodesics. Although the same Crofton's formula is expected to hold in Lorentzian space, the precise form suitable for practical
computation is less clear. We study this problem in the Lorentzian AdS$_3$ and find the area of generic space-like two dimensional surface can be reproduced by the flux of space-like geodesics, which are associated with the entanglement entropies of intervals in the boundary CFT. In our opinions this may be the first step towards understanding the CV conjecture from the perspective of information theory. We analyze this result using some toy model and find it reasonable to interpret the geodesic number contributing to the complexity as counting the entanglement of a state constructed from the path integral with two boundaries. More importantly we realize that these geodesics are circuit-independent and hence contribute as lower bound on the cost of all the circuits.

In section~\ref{ig on e}, we review some basic facts in the integral geometry, particularly the Crofton's formula in the Euclidean space. In section~\ref{ig on m} we figure out the precise form of the Crofton's formula in the Lorentzian AdS$_3$ and show that the area of a generic 2d space-like surface is given by the number of space-like geodesics it intersects. Finally in section~\ref{sec:discussion}, we discuss the possible reason why the complexity can be expressed in terms of the geodesic number.

\section{Integral geometry in Euclidean space}\label{ig on e}
Integral geometry is not a new subject. In fact most of the conclusions we are going to use are probably well known to mathematicians \cite{santal}. It was not until very recently that \cite{czech1} they were introduced to the community of AdS/CFT and provided some interesting new insights. Here we will briefly touch on various useful results for later convenience.

\subsection{Kinematic space and invariant measure}\label{ssbks2}
In integral geometry, geometric objects are expressed in terms of the
integrals of some probe objects, the collection of which forms the so-called kinematic space. In the remainder of the paper, we mostly focus on the kinematic
space of the geodesics in which every point represents a single geodesic. In a symmetric space like $\ads_3$, the
geometry of the kinematic space can be determined entirely on symmetry ground
(as it can be expressed in terms of the coset of symmetry groups).

To see this explicitly, we can pick two points $x_1$ and $x_2$ on the boundary to denote a geodesic
and express the metric as follows
\be \rmd s^2=f_{\mu\nu}(x_1,x_2)\rmd x_1^{\mu}\rmd x_2^{\nu}\quad(\mu,\nu=0,1)\,.\ee
The precise form of $f_{\mu\nu}$ can be obtained using
the following trick in \cite{czech2}: Its transformation under the conformal group is the same as the two-point function of two spin-$1$ currents of conformal dimension $1$. The requirement of conformal invariance then fixes its
form to be
\be f_{\mu\nu}(x_1,x_2)=\frac{4I_{\mu\nu}(x_1-x_2)}{|x_1-x_2|^2},\label{fmunu}\ee
where $4$ is an normalization constant and $I_{\mu\nu}(x) = \eta_{\mu\nu}
- 2x_\mu x_\nu/x^2$.

For simplicity, we will work in the Poincare patch (\ie $x^{0,1}
= t, x$) with metric
\begin{equation}\label{metric}
   \rmd s^2=\frac{\rmd u^2+\rmd z\rmd\bar{z}}{u^2}=\frac{\rmd u^2+\rmd x^2-\rmd t^2}{u^2},
\end{equation}
where we introduce the null coordinates
\begin{equation}
    z=x+t\qquad \bar{z}=x-t\,.
\end{equation}
It is not different to check using eq.\eqref{fmunu} and the precise form
of $I_{\mu\nu}$ that the density for geodesics space becomes
\begin{equation}\label{3adsks}
\rmd s^2=\frac{2\rmd z_1\rmd z_2}{(z_1-z_2)^2}+\frac{2\rmd \bar{z}_{1}\rmd \bar{z}_{2}}{(\bar{z}_{1}-\bar{z}_{2})^2}.
\end{equation}
The same result can be deduced from the second derivatives of the geodesic length as $f_{\mu\nu} (x_1, x_2) =\p_\mu \p_\nu L (x_1, x_2)$, the latter of which in the null coordinates reads ($\xe$ being the cutoff)
\begin{equation}\label{3adsgeol}
  L(z_1,z_2)=\log\lsb\frac {(z_1-z_2)(\bar{z}_1-\bar{z}_2)} \xe\rsb\,.
\end{equation}

The Euclidean $\ads_2$ or rather $\BH^2$ can be regarded as a constant time slice of the $\ads_3$ like the one specified by $z_i=\bar{z}_i$, and we get the kinematic space on the $\BH^2$ with metric
\begin{equation}\label{2adsle}
    \rmd s^2=\frac{4\rmd z_1\rmd z_2}{(z_1-z_2)^2}\,.
\end{equation}

\subsection{Crofton's formula in the Euclidean $\ads_2$ }
In two-dimensional space, the Crofton's formula says that the length of a curve is given by the number of geodesics it meets. As explained above, this number
follows from a volume integral in the kinematic space. We leave the derivation in Euclidean plane in the appendix~\ref{e2}. The more interesting case to us is Euclidean
AdS$_2$ (E$\ads_2$), where the formula takes the following form
\begin{equation}\label{2adscf1}
  L_{\gamma} =\h \int_{M_1\cap L_1\neq0}\frac{2\sigma_{0}(M_1\cap L_1)}{(z_1-z_2)^2}\rmd z_2\wedge \rmd z_2\,,
\end{equation}
where the integration is over all geodesics $L_1$ with nonvanishing intersection
number $\sigma_{0}(M_1\cap L_1)$ with the curve $\xg$. As we can see, the denominator comes from the measure in eq.\er{2adsle}. In general the measure
is given by the second derivatives of the geodesic length and the $\cf$ goes
like
\begin{equation}\label{2adscf2}
  L_{\gamma} =\frac{1}{2}\int_{M_1\cap L_1\neq0} \sigma_{0}(M_1\cap L_1)\frac{\partial^2L(z_1,z_2)}{\partial z_1\partial z_2}\rmd z_1\wedge \rmd z_2.
\end{equation}
Moreover the length of a geodesic is related to the entanglement entropy via the RT formula:
\begin{equation}\label{rt}
 S(z_1,z_2)=\frac{ L(z_1,z_2)}{4G}
\end{equation}
where $S(z_1,z_2)$ is the entanglement entropy of an interval $(z_1,z_2)$ on the boundary. Putting eq.(\ref{rt}) into eq.(\ref{2adscf2}), we get
\begin{equation}\label{2adscf}
  \frac{L_{\gamma}}{4G} =\frac{1}{2}\int_{M_1\cap L_1\neq0} \sigma_{0}(M_1\cap L_1)\frac{\partial^2S(z_1,z_2)}{\partial z_1\partial z_2}\rmd z_1\wedge \rmd z_2.
\end{equation}
So the length $L_{\gamma}$ as a bulk geometric quantity is connected with the entanglement entropy $S(z_1,z_2)$. In fact, given the entanglement entropies of all intervals, one can reconstruct the geometry in the kinematic space and hence the geometry in the bulk. This is a perfect example of notion of ``entanglement=geometry''.

\subsection{The generic Crofton's formula}\label{gcf}
The generic  Crofton's formula first proposed in \cite{santal2} (see also \cite{santal,czech1,huang,convex}) establishes the relationship between $q$-dimensional target object $M_q$ and the sets of $r$-planes (geodesically complete submanifolds) for any constant curvature space. In a $d$ dimensional
Euclidean space, it takes the following form
\begin{equation}\label{gcf1}
\int_{M_q\cap L_r\neq0} \sigma_{q+r-d}(M_q\cap L_r) \epsilon_{\CK }=\frac{O_d...O_{d-r}O_{q+r-d}}{O_r...O_1O_0O_q}\sigma_q(M_q)\,.
\end{equation}
We note that the $r$-planes $L_r$ are unoriented \footnote{This convention is the same as \cite{huang,santal}, but different from \cite{czech1}. The difference leads to a factor $2$ in the Crofton's formula.}. The volume element $\epsilon_{\CK }$ of the kinematic space $\CK$ measures the number density of the $r$-planes. The symbols $\sigma_q(M_q)$ and $\sigma_{q+r-d}(M_q\cap L_r)$ denote the volumes of $n$ and $q+r-d$ dimensional objects, the latter of
which is the cross section between $M_q$ and $L_r$. The numeric factors $O_k$
are the area of $k$ dimensional unit-sphere,
\[O_k=\frac{2\pi^{\frac{k+1}{2}}}{\Gamma(\frac{k+1}{2})}.\]

In this paper, we only consider the kinematic space of geodesics (that is $r=1$) and in this case eq.\eqref{gcf1} reduces to
\be
\label{croftonq1}
\int_{M_{d-1}\cap L_1\neq0} N(M_{d-1}\cap L_1) \epsilon_{\CK }=\frac{O_d}{O_1}\sigma_{d-1}(M_{d-1})\,.
\ee
where $ N(M_{d-1}\cap L_1) \equiv \sigma_{0}(M_{d-1}\cap L_1)$ is the number of intersection points $M_{d-1}\cap L_1$. There are two special cases, $d=2$ and $d=3$, respectively,
\begin{equation}\label{gcf2}
\begin{split}
\left\{
  \begin{array}{ll}
    \sigma_1(M_1)=\frac{1}{2}\int_{M_1\cap L_1\neq0} N(M_1\cap L_1) \epsilon_{\CK }& \hbox{(d=2,r=1,q=1),} \\
    \sigma_2(M_2)=\frac{1}{\pi}\int_{M_2\cap L_1\neq0} N(M_2\cap L_1) \epsilon_{\CK }& \hbox{(d=3,r=1,q=2)}\,.
  \end{array}
\right.
\end{split}
\end{equation}

 One merit of this choice \footnote{Such a choice also extend the Crofton's formula to general Riemannian surface but this is irrelevant in
the current context.} is that the measure is always given by the
second derivative of the lengths of geodesics even in the absence of maximal
symmetry,
\be
\label{generalmeasure}
\xe_\CK = \det \lsb \frac {\p^2 L(\vec x_1, \vec x_2) }{\p \vec x_1 \p \vec x_2}\rsb \prod_{i=1}^{d-1} \rmd x^i_2 \wedge \rmd x^i_1  \,,
\ee
where we still use $L$ to denote the length of a geodesic (even though it no longer has any connection with entanglement entropy) and a geodesic is parameterized
using its coordinates $(\vec x_1, \vec x_2)$ ($x_{1,2}^i,\; i=1,\dots d-1$) of the end points.

\section{Integral geometry in the Lorentzian $\adst$}\label{ig on m}
In this section, we will study the  Crofton's formula on Lorentzian $\adst$. We will stick with the geodesics as the probe (\ie $r=1$) but now they are
no longer restricted to a time slice. One subtlety about the Lorentzian space is that there are three different types of geodesics (time-like, space-like and null). Moreover, the time-like geodesics never hit the boundary twice and they usually have no known information theoretic meaning in the boundary theory (neither are the null geodesics). Fortunately, it turns out that space-like geodesics are enough to see the space-like
2-surface and we have the following Crofton's formula similar to eq.(\ref{gcf2}) \begin{equation}\label{scf}
    \sigma_2(M_2)=\xk\int_{M^2\cap LS_1\neq0} \sigma_{0}(M^2\cap LS_1) \epsilon_{\CK},
\end{equation}
where $LS_1$ denote space-like geodesics and $M_2$ is a space-like 2-surface
and $\xk$ is a numeric factor to be determined later. In the remainder of this section, we will prove this formula.

\subsection{Geodesics}
To prove eq.(\ref{scf}) it is necessary to find all  geodesics passing through a given surface $M_2$. In Poincare coordinate \er{metric}, the parametrization of a geodesic from one boundary point $(z_1,\bar{z}_1,0)$ to another $(z_2,\bar{z}_2,0)$ is \cite{wittendiagram}
\begin{equation}\label{z}
z(\lambda)=\frac{z_1+z_2}{2}+\frac{z_1-z_2}{2}\tanh{\lambda},
\end{equation}
\begin{equation}\label{zbar}
\bar{z}(\lambda)=\frac{\bar{z}_{1}+\bar{z}_{2}}{2}+\frac{\bar{z}_{1}-\bar{z}_{2}}{2}\tanh{\lambda},
\end{equation}
\begin{equation}\label{u}
u(\lambda)=\frac{\sqrt{(z_1-z_2)(\bar{z}_{1}-\bar{z}_{2})}}{2\cosh{\lambda}},
\end{equation}
where $(z,\bar{z},u)$ denote a point along the geodesic and $\lambda$ is a parameter ranging from negative infinity to positive infinity.

A geodesic can provide a nonzero contribution to the integral when it hits  the target object $M_2$, which can be parameterized by the function $\bar{z}=\bar{z}(z,u)$. A bulk point $(z,\bar{z},u)$ on $M_2$ and a boundary point $(z_1,\bar{z}_1,0)$ determine the other
\begin{equation}\label{coor1}
  \begin{split}
    &z_2=z+\frac{u^2}{\bar{z}-\bar{z}_1},\\
   &\bar{z}_2=\bar{z}+\frac{u^2}{z-z_1}.
  \end{split}
\end{equation}
So we can instead use $(z,u,z_1,\bar z_1)$ to denote the geodesics, all of
which are space-like and therefore, we have
\begin{equation}\label{spacelike}
  \begin{split}
    &u^2+(z-z_1)(\bar{z}-\bar{z_1})>0,\\
 \zt{or equivalently}\quad & u^2+(z-x_1-t_1)(\bar{z}-x_1+t_1)>0,
  \end{split}
\end{equation}
Now we consider the target surface. Expressing the derivative of $\bar{z}$ with respect to $z$ and $u$ as
\[A = \frac {\p \bar z}{\p z},\quad B = \frac {\p \bar z}{\p u}\,,\] the space-like constraint of the surface requires that the normal dual vector $(A,-1,B)$ is time-like, that is
$$-4A+B^2<0.$$
\subsection{The Crofton's formula on the $\ads_3$}
We have already calculated the measure of kinematic space of Lorentzian $\ads_3$ in the sec~\ref{ssbks2}, which gives the following integral from eq.\eqref{scf},
\begin{equation}\label{3adsscf1}
\tilde \sigma_2(M_2)=\xk\int \rmd z_2\int \rmd\bar{z}_2\int \rmd z_1\int \rmd\bar{z}_1 \;\frac{\sigma_0(LS_1\cap M_2)}{(z_1-z_2)^2(\bar{z}_1-\bar{z}_2)^2}.
\end{equation}
Here we use the notation $\tilde \sigma_2(M_2)$ to denote the integral and eventually we will see that it is equal to the area of $M_2$. Under coordinate transformation (\ref{coor1}), the right hand side becomes
\begin{equation}\label{3adsscf2}
\begin{split}
 \tilde \sigma_2(M_2)
=\xk\int \rmd z\int \rmd u\int \rmd z_1\int \rmd\bar{z}_{1}\,
\frac{1}{2}\left|\frac{B[u^2-(z-z_1)(\bar{z}-\bar{z}_1)]+2u[A(z-z_1)-(\bar{z}-\bar{z}_1)]}{[u^2+(z-z_1)(\bar{z}-\bar{z_1})]^3}\right|
\end{split}
\end{equation}
The range of the parameters $z,u$ is determined by the surface $M_2$. Given $z$ and $u$, $(z_1, \bar{z}_1, 0)$ take all points satisfying eq.(\ref{spacelike}). It is noteworthy that we compute all the geodesics twice in the case, therefore the eq.(\ref{3adsscf2}) contains a factor $\frac{1}{2}$. The expression in eq.(\ref{3adsscf2}) only depends on the relative position of the bulk and boundary points, and therefore, we introduce the new coordinates $x,t$
\begin{equation}\label{zz}
\begin{split}
 &\hat{z}=z_1-z=x'+t' \\
 &\check{z}=\bar{z}_1-\bar{z}=x'-t'
\end{split}
\end{equation}
which satisfies
$$u^2+x'^2-t'^2>0.$$
After a boost transformation in $x',t'$, one gets
\begin{equation}\label{3adsscf3}
\begin{split}
  &\tilde \sigma_2(M_2)\\
=&\xk\int \rmd z\int \rmd u\int \rmd x'\int \rmd t' \frac{|B(u^2-x'^2+t'^2)-2ut'\sqrt{4A}|}{(u^2+x'^2-t'^2)^3}\\
=&\xk\int \rmd z\int \rmd u\int \rmd x'\int \rmd t' \frac{\sqrt{4A-B^2}|\sinh\xi(u^2-x'^2+t'^2)-2ut'\cosh\xi|}{(u^2+x'^2-t'^2)^3},
\end{split}
\end{equation}
where
\begin{equation}\label{xi}
  \begin{split}
\sinh\xi=\frac{B}{\sqrt{4A-B^2}}\,,\quad\cosh\xi=\frac{\sqrt{4A}}{\sqrt{4A-B^2}}.
  \end{split}
\end{equation}
To take care of the absolute value, we have to go to the angular coordinates
\begin{equation}\label{coor2}
  \begin{split}
&\sin\beta\sinh\alpha=\frac{2ut'}{u^2+x'^2-t'^2},\\
&\sin\beta\cosh\alpha=\frac{u^2-x'^2+t'^2}{u^2+x'^2-t'^2},\\
&\cos\beta=\frac{2ux'}{u^2+x'^2-t'^2}.
  \end{split}
\end{equation}
Physically, we pick three unit vectors $\hat x, \hat t, \hat u$ (vielbeins) along $\p_x, \p_t, \p_u$ and $\xa, \xb$ are the angles between the unit tangent vector $v$ of the geodesic and the vielbeins. More precisely, $\xb$ is the angle with $\hat x$ and $\xa$ is the (hyperbolic) angle between $v-(v \cdot \hat x) \hat x$ and $\hat u$. The Jacobian then reads
\begin{equation}\label{J2}
  \left|\frac{\partial(x',t')}{\partial(\alpha,\beta)}\right|=\frac{u^2|\sin\beta|}{(1+\cosh\alpha\sin\beta)^2}\,,
\end{equation}
and eq.(\ref{3adsscf3}) becomes
\begin{equation}\label{3adsscf4}
\begin{split}
&\tilde \sigma_2(M_2)\\
=&\xk\int \rmd z\int \rmd u\int_0^{2\pi} \rmd\beta\int_{-\infty}^{+\infty}\rmd\alpha \frac{\sqrt{4A-B^2}}{4 u^2}|\sin\beta||\sin\beta\sinh(\xi-\alpha)|\\
=&\xk\int \rmd z\int \rmd u \frac{\sqrt{4A-B^2}}{4u^2}(\cosh\alpha|^{-\frac{1}{\chi}}_0+\cosh\alpha|^{+\frac{1}{\chi}}_0)\\
=&\xk\int\rmd z\int \rmd u \frac{\sqrt{4A-B^2}}{2u^2}\lb \cosh\frac{1}{\chi}-1\rb,
\end{split}
\end{equation}
where $\chi$ is a cutoff for $\alpha$. We note that $|\sin \xb| \rmd \xb \rmd \xa$ is the volume element of the solid angle and $\sin \xb \sinh(\xi-\alpha)$ is the inner product between $v$ and the normal vector $\hat n$ of the surface, which implies this integral should be independent of the $\hat n$. Practically the parameter $\xi$ drops out after a shift in $\xa$ (which is equivalent to choosing new vielbeins with the normal vector as $\hat t$).

As a quick consistent check, we can perform the same integral in the Euclidean space. With $\hat n$ being one the of axes, the integral ($\xt$ being the angle with $\hat n$)
\be
\frac 1 2 \int |\cos \xt| \rmd \xO_{d-1} = \int_0^{\frac \pi 2} \cos \xt \sin^{d-2} \xt \rmd \xt \rmd \xO_{d-2} = \frac {O_d}{O_1} \,,
\ee
gives precisely the numeric factor on the right hand side of eq.\eqref{croftonq1}.

We can now compare the final result \eqref{3adsscf4} with the area of $M_2$. From the induced line element
\begin{equation}\label{induced line element}
  \rmd s^2=\frac{A\rmd z^2+B\rmd z\rmd u+\rmd u^2}{u^2},
\end{equation}
one may get
\begin{equation}\label{area}
\sigma_2(M_2)=\int \rmd z\int \rmd u \frac{\sqrt{4A-B^2}}{2u^2}\,,
\end{equation}
which agrees with (\ref{3adsscf4}) up to an infinite factor, which is canceled by $\xk$
\be
\label{infinitefactor}
\xk^{-1} = \h \int_0^{2\pi} \rmd\beta\int_{-\infty}^{+\infty}\rmd\alpha \sin^2\beta|\sinh(\alpha)| = \cosh\frac{1}{\chi}-1\,.
\ee
As a result, with the CV assumption complexity (whether it is that of a pure state or the reduced density matrix of a subregion \cite{Alishahiha:2015rta}) can be expressed in terms of the number of geodesics.

\section{Discussions}
\label{sec:discussion}
We examined the precise form of the Crofton's formula in the Lorentzian AdS$_3$ and showed that the area of a space-like two dimensional surface is given by the flux of space-like geodesics. Based on the validity of the Crofton's formula in general Euclidean AdS, we expect the same conclusion to hold for space-like codimension-one surfaces in higher dimensional Lorentzian asymptotically AdS spaces. In AdS$_3$, the geodesics have entropic interpretation and hence it is tempting to think that this conclusion may provide an information theoretic explanation of the CV conjecture. We would like to share some of our observations in that regard, leaving the more complete analysis to future study. For simplicity, we only consider the complexity of a pure state.

It was proposed in \cite{huang} that one can heuristically associate every geodesic with a Bell pair located at the two end points on the boundary. By no means this naive picture captures all the physics as the entanglement structure is not entirely bipartite. It does however offer a very nice interpretation of the entanglement entropy (of a single interval) as counting the number of Bell pairs crossing the entangling surface. Moreover, the length of a convex bulk curve (\ie differential entropy) can also be understood as the amount of long-range entanglement in this framework.

\begin{figure}[htbp]
  \centering
  \includegraphics[width=8cm]{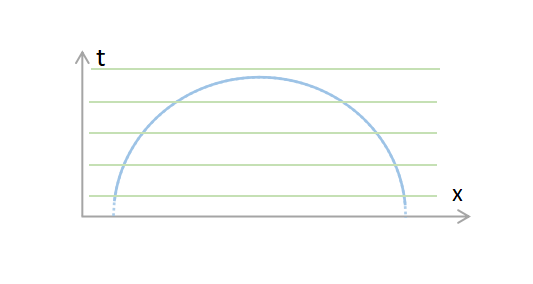}\\
  \caption{Contribution from a certain geodesic to complexity. Along the RG flow (depicted as horizontal straight-lines), a certain number of isometries (reshuffling operation) contribute to the complexity before the disentangler (removing operation) gets rid of the geodesic entirely. For simplicity we can consider the discrete case and assume that the number of sites is reduced by half at each step of RG and precisely one isometry is counted. A geodesic with $|x_2- x_1| = \ell$ can undergo $\log_2 \ell \sim \log \ell$ steps and hence the total number of gates is roughly the length of the geodesic.}\label{complexityfig1}
\end{figure}

We find this picture also very illuminating in the current context and hence decide to stay with it in subsequent discussions despite its apparent flaw. To avoid the issue of Bell pairs, one can simply take the geodesic density as a type of measure of the two-point entanglement. Based on this picture, it was pointed out \cite{huang} that under renormalization group flow, the short-range entanglement is removed while the long-range entanglement is reshuffled to shorter scales. These two operations are the ``geodesic'' versions of disentangler and isometry in MERA. Complexity counts the total number of these two operations. From the perspective of one geodesic, it contributes one removing (disentangler) operation or one reshuffling (isometry) operation for each step of RG, with the total number proportional to the length of the geodesic (see Fig.\ref{complexityfig1}). Consequently, the complexity is given by counting the total number of geodesics weighted by the length of each, which is precisely the area of the codimension-one bulk surface according to the Crofton's formula applied to the constant time slice $\BH^2$ alone \cite{Abt:2018ywl}.

The Crofton's formula in AdS$_3$ can reproduce not only the complexity (area of the corresponding optimized surface) of a state but also the cost of a non-optimized circuit, which may help to understand the CV conjecture from the viewpoint of information theory. It is however very unfortunate that henceforth we have to restrict ourselves to the Euclidean AdS. It is unlikely that the area of a generic codimension-one surface gives the cost of the circuit in the Lorentzian case. Obviously that is not the case for a time-like or null-like surface. Moreover, in the Lorentzian case it is usually the maximal surface that corresponds to the optimized circuit and gives the complexity. The infinite factor \er{infinitefactor} between the flux of geodesics and the area makes it difficult to connect the former to any information theoretic interpretation. Nevertheless, we still hope the subsequent discussions in the Euclidean case may shed some lights on the Lorentzian problem that we eventually have to tackle.

\begin{figure}[tbp]
\begin{minipage}[t]{0.5\textwidth}
\includegraphics[scale=0.7]{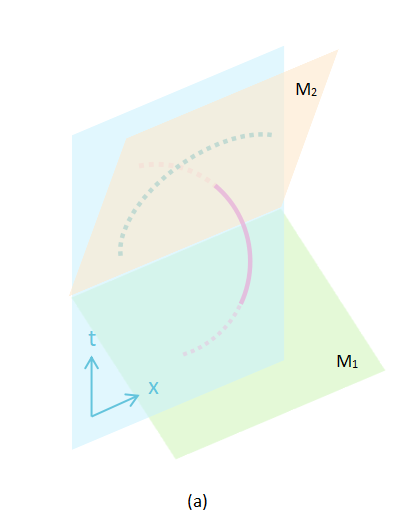}
\end{minipage}
\begin{minipage}[t]{0.5\textwidth}
\includegraphics[scale=0.75]{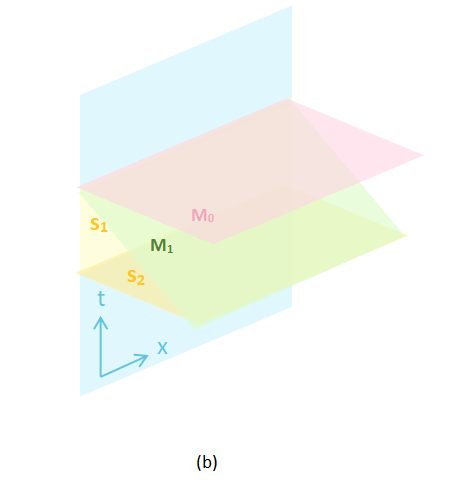}
\end{minipage}
\caption{(a) A geodesic (pink) with two end points on different sides of $t=0$ always contributes to the cost of the circuit ($M_1, M_2$) while the one (blue) with both end points on the same side does not. We also note that the number density of geodesics can be obtained from the second derivative of the lengths of geodesics between two surfaces $M_1 ,M_2$. (b) A tilted plane is a less optimized circuit and it can be decomposed as the combination of an optimized circuit $s_1$ at time $t=-\xd t$ and a circuit $s_2$ for the time evolution back to $t=0$.}\label{complexityfig2}
\end{figure}

In the Euclidean case, the correspondence between circuit and co-dimension one surface, combined with the CV conjecture implies that the cost of a circuit is measured by the area of the surface, which in turn follows from the flux of geodesics via Crofton's formula. It is not clear to us why the number density of geodesic actually accounts for the cost. The good news is that the former does follow from entanglement entropy associated with the quantum circuit, which was computed in \cite{Caputa:2018xuf}. The conformal factors $e^{2\phi}$ at the end points provide corrections $c/6\, \phi$ to the entanglement entropy of a single interval
\be
S(x_1,x_2) = \frac c 3 \log |x_1 - x_2| + \frac c 6 \lsb \phi (x_1) + \phi (x_2)\rsb\,,
\ee
where $x_1, x_2$ are the coordinates of the end points. Let us consider a simple example of the entanglement entropy at $t = \mu$, \ie that of the excited state corresponding to the bulk curve on the $t = 0$ time slice specified by $u = \mu$. The change in the conformal factor from $e^{2\phi(t=\xe, x)} = \frac 1 {\xe^2}$ to $e^{2\phi(t=\mu, x)} = \frac 1 {\mu^2}$ implies that the entanglement entropy is given by $\frac c 3 \log \lb \ell/\mu \rb$ for an interval of length $\ell$, which agrees with the length of a geodesic on the new cutoff surface. We would like to remind the reader that the number density is obtained from the length on geodesics ending on new surface $M[\xS(0)]$ (see \eg Fig.\ref{complexityfig2}(a)), which is guaranteed by the nontrivial fact that the measure in the kinematic space of geodesics always follows from eq.\eqref{generalmeasure} even in a general space without any symmetry. The same conclusion does not necessarily hold for the probes of higher dimensions.

It is very clear that the flux of geodesics depends on the circuit/surface. Graphically, we know that the optimized surface (for vacuum state at $t=0$) receives no contribution from geodesics with both end points in the $t < 0$ region. Instead, every circuit receives contribution from the geodesics connecting the $t>0$ and $t<0$ regions (see Fig.\ref{complexityfig2}(a)). Such a circuit independent contribution serves as the lower bound of the cost, which is saturated by the optimized circuit (corresponding to the $t=0$ slice in the bulk, henceforth $M_0$).

We would like to explore the physical meaning of such a contribution. In \cite{Takayanagi:2018pml}, it is shown that the relevant geodesics come from the entanglement entropy between subsystem $AA'$ and $BB'$, with $A,B$ being subsystems on a slice in the $t<0$ region while $A',B'$ being subsystems on a slice in the $t > 0$ region. The quantum state of the total system is obtained from the mapping given by the path integral with the two slices as the boundaries. More precisely, a mapping like $\st{i}A_{ij}\br j|$ leads to an in general entangled state $\st{i}\st{j} A_{ij}$ by turning bras into kets. Such a practice is common in the study of tensor network. The identity map $\st{i}\br i|$ becomes Bell pairs (more precisely a maximally entangled state) after the move.

\begin{figure}[htbp]
  \centering
  \includegraphics[width=8cm]{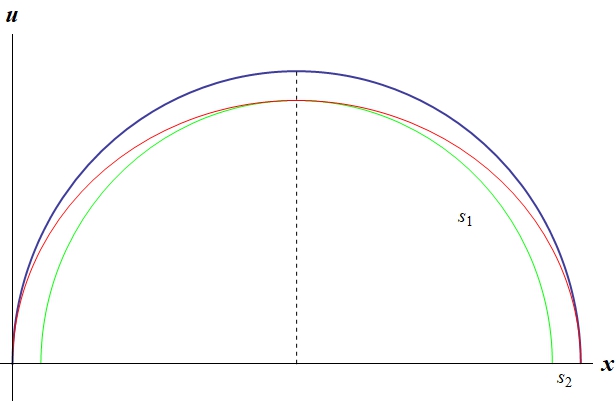}\\
  \caption{To get a better look at the circuit-dependent flux, we go down to EAdS$_2$. The geometry in the Poincare patch discussed earlier is realized in the region to the right of the dashed line. The optimized circuit is a semi-circle (blue) while the less optimized one is the red curve sharing the same boundary. The green semi-circle (optimized circuit) serves as $s_1$. Geodesics that hit the red but not the blue one are those between $s_1$ and $s_2$.}\label{complexityfig3}
\end{figure}

For better demonstration, let us assume the system is discrete and the state takes the form of a tensor network obtained from RG flow (realized as mappings between various Hilbert spaces with different dimensions, see \eg \cite{vidal2009entanglement} for a review). The PI integral then becomes the mapping between two slices $t_2=-t_1>0$, which is given by $W^\dagger(0, t_1) W(0, t_1)$ with $W(0, t_1)$ being the mapping of RG flow. It is known that such a product should be proportional to identity $W^\dagger(0, t_1) W(0, t_1) = \xa I$ \footnote{$W$ is a map between Hilbert spaces at different energy scales and therefore we need $\xa$ to take into account the difference in dimensions.} and hence there are Bell pairs between the subsystems at $t_1$ ($AB$) and $t_2$ ($A'B'$) whose total number gives the entanglement entropy between the two subsystems.

The next step is to consider states constructed from two arbitrary slices $t_1, t_2$ (assuming $0<t_2 < |t_1|$). The unitary $W(0,t_1)$ can be broken up as $W(0,-t_2)W(-t_2, t_1)$. The mapping from $t_1$ to $t_2$ then becomes
\[W^\dagger(0, -t_2) W(0, t_1) = \xa I_{-t_2} \, W(-t_2, t_1)\,.\]
The reduced density matrix for slice $t_1$ is obtained by tracing out the system at $-t_2$
\[\rho_{t_1} \sim W^\dagger (-t_2, t_1) W(-t_2, t_1)  \sim I_{t_1}\,,\]
which is equal to the identity operator again after appropriate normalization and hence the total entanglement entropy is determined by the size of the Hilbert space at the scale of $t_1$. The interesting part is that the specific forms of $W(0, t_1)$'s, which lead to new quantum circuits do not affect the total number of Bell pairs \ie the entanglement entropy. To see that we first rewrite the new circuit $W'(0, t_1)$ as the product of (see Fig.~\ref{complexityfig2}(b))
\[W'(0, t_1) = U(\xd t) U^\dagger(\xd t) W(0, t_1) U(\xd t)\,,\]
where $U(\xd t)$ is the time translation by the amount $\xd t$. The mapping to consider is
\[W^\dagger(0, -t_2) W'(0, t_1) = W(-t_2, t_1) U(\xd t)\,,\]
(or $W^\dagger(t_1, -t_2) U(\xd t)$ if $t_2 > |t_1|$) and one can then use the same argument to show the invariance of the entanglement between $t_1$ and $t_2$. The slice $t_2$ is on the surface $M_0$ while the slice $t_1$ is on $M$. In the heuristic picture discussed above a Bell pair turns into a geodesic in the continuous limit and the entanglement is measured by the geodesics with one point on $M_0$ and the other on $M_1$. In fact since $M_1$ is closed and shares the boundary with $M_0$, every geodesic going through $M_0$ must also hit $M_1$. What we learn from the circuit point of view is that the flux of these geodesics corresponds to the circuit-independent entanglement. For comparison, we can also take a look at the circuit-dependent contribution. As we can see from Fig.~\ref{complexityfig3}, the extra flux follows from the geodesics between $s_2$ and $s_1$, which is the consequence of the additional piece $s_2$ corresponding to the circuit $U(\xd t)$.

Despite only a hand-waving argument, it does give a reasonable picture in which the flux of geodesics going through $M_0$ measures the entanglement between the different Euclidean times (or energy scales in the RG sense) and provides the circuit-independent contribution to the cost. This lower bound is saturated when other circuit-dependent contributions all drop out \ie when surface is $M_0$. 

\section*{Acknowledgments}
XH is supported by the NWU Starting Grant No.0115/338050048 and the Double First-class University Construction Project of Northwest University.

\appendix
\section{The Crofton's formula on the Euclidean plane}\label{e2}
In this section we prove the Crofton's formula on the Euclidean plane $E_2$
(with
the metric $\rmd s^2=\rmd x^2+\rmd y^2$), which measures a smooth convex closed curve $\gamma$ by a set of geodesics and verify that its integrand is invariant under isometry.
\begin{figure}[htbp]
  \centering
  \includegraphics[width=8cm]{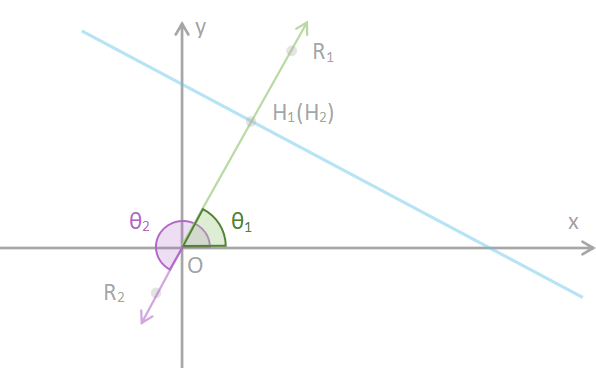}\\
  \caption{If the direction of $OH$ is the same as that of $OR$, $p=\overline{OH}$; otherwise $p=-\overline{OH}$. We can define the positive orientation as the direction of angle $\theta+\frac{\pi}{2}$ with the axis Ox. In this convention, there is no degeneracy in the parametrization of $(p, \xt)$. }\label{orientation}
\end{figure}

We start with a geodesic in $E_2$. Shoot a ray $OR$ from the origin which is perpendicular to the given geodesic at point $H$. Let $\theta$ be the angle between the ray $OR$ and the $x$-axis, and let $p$ be the distance of the line segment $OH$ as in figure \ref{orientation}. The equation of geodesic is then given by
\begin{equation}\label{2epgeo1}
x\cos\theta+y\sin\theta-p=0.
\end{equation}
A geodesic specified by $\theta$ and $p$ is identified with the other parameterized by $\theta+\pi$ and $-p$, as we can see in figure \ref{orientation}. With
the introduction of the orientation to the geodesics, the degeneracy is lifted.

For convenience, let us consider a smooth convex closed curve $\gamma$. For a given angle $\theta$, there are two straight-lines tangent to the curve $\gamma$ and  we make such a convention that $p(\theta)$ is the larger of the two. In fact, $p(\theta+\pi)$ will correspond to the other one. So $p$ is a single-value function of $\theta$ with period $2\pi$. All such geodesics forming an envelope of $\xg$ and their equations \eqref{2epgeo1} can be expressed in terms of a single implicit function as $F(x,y,p(\theta),\theta)=0$. According to the envelope theorem, $F=0$ and $\partial_{\theta}F=0$ determine the curve
\begin{equation}\label{2epcurve1}
\begin{split}
   &x=p\cos\theta-\frac{\rmd p}{\rmd\theta}\sin\theta,\\
   &y=p\sin\theta+\frac{\rmd p}{\rmd\theta}\cos\theta.
\end{split}
\end{equation}

The conditions for $\xg$ to be smooth, convex and closed implies that $p+\frac {\rmd^2 p}{\rmd \xt^2}>0$ and $\frac{\rmd p}{\rmd\theta}|_{\theta}=\frac{\rmd p}{\rmd\theta}|_{\theta+2\pi}$. Now we can compute the length of the curve $\gamma$ as
\begin{equation}\label{2eplength1}
  L_{\gamma}=\oint \rmd s=\int_{0}^{2\pi}\rmd\theta\: |p+\frac {\rmd^2 p}{\rmd \xt^2}|=\int_{0}^{2\pi}\rmd\theta\:p\,,
\end{equation}
which can be rewritten as
\begin{equation}\label{1cs3}
\begin{split}
  L_{\gamma}=&\frac{1}{2}\int_{0}^{2\pi}\rmd\theta\:p(\theta)+\frac{1}{2}\int_{0}^{2\pi}\rmd\theta\:p(\theta+\pi) \\
    =& \frac{1}{2}\int_{0}^{2\pi}\rmd\theta \int_{-p(\theta+\pi)}^{p(\theta)}\rmd p=\frac{1}{4} \int_{0}^{2\pi}\rmd\theta \int_{-\infty}^{\infty}\rmd p \;\sigma_0(\gamma\cap L_1)\\
    =&\frac{1}{2} \int_{0}^{\pi}\rmd\theta \int_{-\infty}^{\infty}\rmd p \;\sigma_0(\gamma\cap L_1),
\end{split}
\end{equation}
where $\sigma_0(\gamma\cap L_1)$ denotes the number of intersections between
$\xg$ and $L_1$ parameterized by $p, \theta$. Introducing a differential form $\epsilon_\CK=\rmd p\wedge \rmd\theta$, the equation (\ref{1cs3}) becomes
\begin{equation}\label{cf1}
 L_\gamma =\frac{1}{2}\int_{\gamma\cap L_1\neq\varnothing}\;\sigma_0(\gamma\cap L_1)\epsilon_\CK\,,
\end{equation}
which is the Crofton's formula in $\BE_2$.

Let us derive  the measure of the geodesics on the Euclidean plane from symmetry considerations. The measure $f(p, \theta)\rmd p\wedge \rmd\theta$ shall be
invariant under the isometry transformation
\begin{equation}\label{2episo}
 \left( \begin{array}{c}x \\  y \\ \end{array} \right) =\left(\begin{array}{cc} \cos\phi & -\sin\phi \\ \sin\phi & \cos\phi\\\end{array}\right) \left( \begin{array}{c} x' \\  y' \\ \end{array}\right)+ \left( \begin{array}{c} a \\  b \\ \end{array} \right),
\end{equation}
where $\phi$ is the rotation angle and $a,b$ describe the translation. Plugging (\ref{2episo}) into (\ref{2epgeo1}), the relation between new parameters $p', \theta'$ and old ones $p,\theta$ is
\begin{equation}\label{2epod}
  \theta'=\theta-\phi;\qquad p'=p-a\cos\theta-b\sin\theta.
\end{equation}
Symmetry then requires the new measure $f(p',\xt') \rmd p' \wedge \rmd \xt'$ to agree with the original one. It's easy to check $\rmd p\wedge \rmd\theta=\rmd p'\wedge \rmd\theta'$. The invariance of measure implies for any set X
\begin{equation}
  \int_X\; f(p,\theta) \rmd p\wedge \rmd\theta = \int_X\; f(p',\theta') \rmd p'\wedge \rmd\theta'
  =\int_X \;f(p-a\cos\theta-b\sin\theta, \theta-\phi)\rmd p\wedge \rmd\theta\,,
\end{equation}
which forces $f(p,\theta)$ to be a constant.

\bibliography{bio}

\end{document}